\documentclass[pre,aps,twocolumn]{revtex4}
\usepackage{graphicx}
\def\pt{$p_T$ }
\def\iaa{$I_{AA}$ }
\def\46{$4<p_T^{\rm trig}<6$ }
\begin{document}

\title{Recombination of Shower Partons at High
$p_T$ in Heavy-Ion Collisions}
\author{Rudolph C. Hwa$^1$ and  C.\ B.\ Yang$^{1,2}$}
\affiliation{$^1$Institute of Theoretical Science and Department of
Physics\\ University of Oregon, Eugene, OR 97403-5203, USA\\
\bigskip
$^2$Institute of Particle Physics, Hua-Zhong Normal University,
Wuhan 430079, P.\ R.\ China}

\begin{abstract}
A formalism for hadron production at high \pt in
heavy-ion collisions has been developed such that all partons
hadronize by recombination. The fragmentation of a hard parton is
accounted for by the recombination of shower partons that it creates.
Such shower partons can also recombine with the thermal partons to
form particles that dominate over all other possible modes of
hadronization in the $3<p_T<8$ GeV range. The results for the high
\pt spectra of pion, kaon, and proton agree well with experiments.
Energy loss of partons in the dense medium is taken into account on
the average by an effective parameter by fitting data, and is found
to be universal independent of the type of particles produced, as it
should. Due to the recombination of thermal and shower partons, the
structure of jets produced in nuclear collisions is different from
that in $pp$ collisions. The consequence on same-side correlations is
discussed.
\end{abstract}

\maketitle

\section{Introduction} In the production of hadrons at high $p_T$ in
heavy-ion collisions there are by now three theoretical collaborations
that have shown the importance of quark recombination
\cite{hy,gr,fr}.  While there are some differences among the three
approaches, they all agree in the basics and in the successful
interpretation of the experimental data, among which the most
outstanding ones are the $p/\pi$ ratio being around 1 in the $3 < p_T
< 4$ GeV range \cite{ppi} and the scaling law of elliptic flow in the
number of constituents \cite{v2,sor}.  The differences concern mostly
with the treatment of hadronization in the $4 <p_T<8$ GeV range.  In
this paper we show how the particles produced in that range arise from
the recombination of the thermal partons and the shower partons
created by hard partons at high $p_T$.  The determination of the
shower parton distributions (SPD) has recently been achieved by
studying the fragmentation functions (FF) in the framework of the
recombination model \cite{hy2}.  This paper contains the first
application of the SPD outside the realm of parton fragmentation.  As
a consequence we find a new component that stands between the
recombination of soft thermal partons at lower
$p_T$ and the fragmentation of hard partons at higher $p_T$
\cite{fr}, and is also different from the direct recombination of soft
and hard partons \cite{gr}.

In \cite{hy} we have avoided the need to specify the origin of the
partons before recombination in order to be independent of the models
describing the early and intermediate phases of the evolution in
heavy-ion collisions.  We use the measured pion spectra to infer the
quark and antiquark distributions just before hadronization, and then
use those distributions to determine the proton spectrum.  In that way
the calculated $p/\pi$ ratio is a direct consequence of the
recombination model with essentially no dependence on the other
aspects of the dynamics such as the separation into soft and hard
components.  In this paper we do enter into the origins of the
partons.  The main difference between our approach and the
conventional treatment of particle production at high $p_T$ is that we
do not use FF to represent the hadronization of hard partons produced
at higher $p_T$.  In our view all hadrons are produced by
recombination at any $p_T$. FF's are phenomenological functions that
do not specify the hadronization mechanism.  Although string models
can be useful in the description of hadronization of a pair of $q$ and
$\bar{q}$ receding from each other after being created in vacuum in,
for example, $e^+e^-$ annihilation, they cannot be applied to
heavy-ion collisions where the abundance of color charges renders
invalid any notion of stretched color flux tubes between pairs of
partons produced in hard collisions \cite{ht}.  In the absence of such
models to give a conceptual basis for fragmentation, it is necessary
to find a meaningful hadronization scheme for hard partons at high
$p_T$.

Our view is that a hard parton creates a shower of partons that
recombine subsequently to form hadrons.  Although the branching
process of gluon radiation and pair creation that eventually lead to
shower partons at low virtuality cannot be calculated, the
distributions of the shower partons can nevertheless be determined
from known FF's in the framework of the recombination model, analogous
to how the $q$ and $\bar{q}$ distributions are determined from the
experimental distribution
$dN/p_Tdp_T$ of pions, as done in \cite{hy}.  In \cite{hy2} a variety
of SPD's are given as functions of the momentum fractions of the
shower partons.  In principle, there should be dependences on $Q^2$;
however, for use in heavy-ion collisions at RHIC, the SPD's given in
\cite{hy2} for $Q^2 = 100$ GeV$^2$ is adequate, since the dependence
on $Q^2$ is not severe.

With the SPD's at hand, we can now consider their role in heavy-ion
collisions.  Hard scattering of partons gives rise to partons with
high $p_T$, which undergo energy degradation as they traverse the
dense medium \cite{wa,gy}.  Instead of tracing the spatial coordinates
of the hard partons, we shall use an effective parameter $\xi$ to
describe the average fraction of partons that escape the dense medium
and are able to hadronize outside.  The value of $\xi$ will be
determined phenomenologically, and will represent an independent check
on the degree of jet quenching
\cite{jq}.  A more important issue is the $p_T$ dependences of the
hadrons detected and the origins of the partons that contribute to the
formation of those hadrons at different regions of $p_T$.  The
recombination of thermal and shower partons will be shown to be
important with the consequence that the structure of jets produced in
heavy-ion collisions is different from that produced in $pp$
collisions.  That difference will manifest itself in the correlation
between particles in the same jet.

The recombination of two shower partons in the same jet is the same as
the usual fragmentation of hard partons and becomes important at
higher $p_T$.  The recombination of two shower partons arising from
two neighboring hard partons is also possible, but will not become
important until the collision energy is very high, e.g., at LHC.

In what follows the production of mesons (pions and kaons) will be
considered first, and then the baryons (specifically proton).
Same-side correlation will be discussed, but only qualitatively to
explain some apparent puzzle in the data.

\section{Meson Production by Recombination}

\subsection{General considerations}

In an overview of a heavy-ion collision we can divide the process into two
stages: (1) the initial evolutionary phase, and (2) the final hadronization
phase. Our concern in this paper is mainly on the latter phase, although what
partons recombine depends on the former. A proper formulation of the formation
of a dense medium must be done in full 3D-space and time, but to describe
hadronization at large transverse momentum can be much simpler.
Since the spatial volume in which recombination can occur is small,
any two partons that are not collinear cannot
recombine. Thus collinear parton momenta in 1D are all that we need
to consider for the
production of a particle with a particular momentum $\vec p$. The
spatial extension
along that direction is also constrained to the size of the hadron,
so an integration
of the remaining spatial variable results in a momentum-space
description in which the
parton momenta should reflect the wave function in momentum-space
representation  of
the produced hadron. We can therefore formulate the recombination
process in the 1D
momentum space only.  Starting with $\vec p$ we investigate the
partons that can recombine to form the particle at that $\vec p$.
Of course, by not starting with a 3D space from the outset, we cannot
calculate the density and expansion properties of the soft thermal partons,
which will have to be introduced phenomenologically.
A more elaborate formulation can be given in terms of Wigner
functions in 6D coordinate-momentum space \cite{gr,fr}. Such formulations
do provide a more complete picture of the collision process with the added
advantage of being able to specify the soft component dynamically.
It should be emphasized that whereas the recombination
process is considered by us  in 1D, it does not mean that the formalism is
insensitive to the realistic problem of
heavy-ion collisions and the 3D nature of the colliding nuclei. We
shall return to this point when we discuss centrality dependence. As far as
the hadronization part of the problem is concerned, our formulation in the 1D
momentum space is consistent with the result of the formulation in 6D
coordinate-momentum space after integration over the other 5 variables.

In our present problem of heavy-ion collisions we shall consider
particle production in
the transverse plane at rapidity $y=0$ only.
   Although the transverse plane is 2-dimensional, we shall
consider only the direction in which a hadron is detected so that all
partons relevant to the formation of such a particle move in the same
direction.  Thus it is unnecessary to carry the subscript $T$ on all
our momentum labels to denote ``transverse.''   The invariant phase
space element is therefore
$dp/p^0$, which we shall approximate by
$dp/p$ for relativistic particles.

In the recombination model \cite{hw} the invariant inclusive
distribution for a produced meson with momentum $p$ is
\begin{eqnarray} p{dN_M  \over  dp} = \int {dp_1 \over  p_1}{dp_2
\over p_2}F_{q\bar q'} (p_1, p_2) R_M(p_1, p_2, p)\ ,
\label{1}
\end{eqnarray} where $F_{q\bar q'} (p_1, p_2)$ is the joint
distribution of a quark $q$ at $p_1$ and an antiquark $\bar{q}'$ at
$p_2$, and $R_M(p_1, p_2, p)$ is the recombination function (RF) for
$q\bar q' \to M$.  For central collisions it is immaterial which direction $\vec{p}$ points.
Although we only consider $\vec{p}$ in the transverse plane here, Eq.\
(\ref{1}) has been used in the longitudinal direction in
\cite{hw,hy3}, where the RF is specified. It is
\begin{eqnarray} R_M(p_1, p_2, p)& = &{1  \over  B(a + 1, b+1)}
\left({p_1  \over p}
\right)^{a+1}\left({p_2  \over p} \right)^{b+1}\nonumber\\
&& \cdot\delta\left({p_1
\over p}+ {p_2  \over p} - 1\right),
\label{2}
\end{eqnarray}
where $B(m,n)$ is the beta function.  For pion it is
shown in
\cite{hy3} from the analysis of Drell-Yan production data in
pion-initiated process \cite{su} in the framework of the valon model
\cite{hw} that $a = b = 0$.  Thus we have explicitly
\begin{eqnarray} R_{\pi}(p_1, p_2, p) ={p_1p_2  \over p^2}
\delta\left({p_1
\over p}+ {p_2  \over p} - 1\right),
\label{3}
\end{eqnarray} The statistical factor for the recombination process is
1 \cite{hy}. For kaon it follows from the constituent quark masses of
the valons and kaon-initiated inclusive production
\cite{hy3} that
$a=1$ and $b = 2$.

The $\delta$-function in Eq.\ (\ref{2}) guarantees the conservation of
momentum in the recombination process.  $R_M$ is an invariant
distribution that is related to the non-invariant probability density
$G_M (y_1, y_2)$ of finding the two valons in
$M$ with momentum fractions $y_1$ and $y_2$ by
\begin{eqnarray} R_M(p_1, p_2, p) =y_1y_2  G_M (y_1 y_2 ),
\hspace{1cm} y_i = p_i/p\ .
\label{4}
\end{eqnarray} Thus for pion $G_{\pi}(y_1 y_2 )$ is a constant, apart
from the
$\delta$-function.  It means that in the momentum space the wave
function of pion in terms of the valons is very broad, corresponding
to the pion being a tightly bound state of its constituent quarks.
That is not the case  for kaon.  The broadness of  $G_{\pi}(x_1 x_2 )$
is an important reason why the thermal-shower recombination in the
formation of pions makes a dominant contribution in the intermediate
$p_T$ range, as we shall see below.

It should be noted that $R_{\pi}$ given in Eq.\ (\ref{3}) differs from
the RF given in \cite{hy}, where our
formulation of the high-$p_T$ problem is in the 2D transverse plane.
Even the dimension of $R_{\pi}$ in \cite{hy} is equivalently
$p^{-2}_T$, whereas it is dimensionless in Eq.\ (\ref{3}).  Thus
caution should be exercised in relating our treatment here to that in
\cite{hy}.

It is important to recognize that the RF describes the probability of
recombination of quarks (and antiquarks).  In the recombination model
the gluons are not regarded as partons that can directly hadronize
\cite{hw,hy3}.  They hadronize through the
$q$ and $\bar{q}$ channels by conversion to $q\bar{q}$ pairs. Thus
$F_{q\bar q'}$ in Eq.\ (\ref{1}) must include all the
$q$ and $\bar q'$ generated by gluon conversion for the purpose of
hadronization through the use of $R_M$ in Eq.\ (\ref{1}).  The sea is
therefore saturated by complete conversion from the gluons, a
procedure that leads to the correct inclusive cross section for the
hadronic production of pions, both in normalization \cite{hw} and in
the momentum distribution
\cite{hw,hy3}.  The saturation of the sea of $q$ and $\bar q'$ makes
possible that the momenta of the gluons (which carry roughly half of
the momentum of each nucleon) can be properly accounted for in the
produced hadrons in any hadronic or nuclear collision.  This procedure
of saturating the sea is also followed implicitly in
\cite{hy2} for the determination of the shower partons in the
fragmentation of an initiating parton.  We also note that the
conversion of gluons
to quark pairs increases the number of degrees of freedom of all the
partons in the
medium and thereby overcomes the decrease of entropy that one might
otherwise conclude
by focussing on individual recombination processes.

Lastly, we mention that soft gluon emission and absorption are always
possible in a
recombination process; in fact, that is how color mutation can take
place for a $q\bar q'$
pair to become colorless, while the partons dress themselves to
become valons before
forming a meson. Such soft processes do not change the momenta of
valons and therefore
do not influence the momentum consideration in Eq.\ (\ref{2}). We do
not consider
quark-antiquark-gluon recombination with large momentum fraction for
the gluon because
the valon representation is complete. That is precisely the reason
why the valon model
was constructed in the beginning to describe hadron structure on the
one hand as well
as the recombination function (for the time-reversed process) on the
other \cite{hw}.

\subsection{Pion distribution}

  Restricting our attention to pion production for now, we obtain from
Eqs.\ (\ref{1}) and (\ref{3})
\begin{eqnarray} {dN_{\pi}  \over  pdp} =  {1 \over p^3} \int^p_0
dp_1F_{q\bar q'}(p_1, p-p_1)  ,
\label{5}
\end{eqnarray} which clearly exhibits the simple dependence of the
pion spectrum on the momentum distributions of the $q$ and
$\bar q'$ that recombine.
It should be recognized that, since we work in 1D, $dN_{\pi}/pdp$ in
Eq.\ (\ref{5}) is actually the number density of pions in
$p_Tdp_Tdyd\phi$ evaluated at $y=0$. With the assumption that it is
independent of
$\phi$ in central collisions, what we denote as $dN_{\pi}/pdp$ is
equivalent to the experimental $dN/(2\pi p_Tdp_T)$, where the number
$N$ refers to the integrated result over all $\phi$ so that when
divided by $2\pi$ the average density in 2D is the same as what we
calculate in 1D.

  Note that in Eq.\ (\ref{5}) $p_1$ is integrated over
a wide range due to the broadness of the RF. There are two components
of parton sources that contribute to
$F_{q\bar q'}$.  One is thermal
$(\cal{ T})$ and the other shower $(\cal{S})$.  Before giving the
specifics of what they are, let us first express
$F_{q\bar q'}$ in terms of them in a schematic way:
\begin{eqnarray} F_{q\bar q'} = {\cal TT} + {\cal TS} + ({\cal SS})_1 +
({\cal SS})_2 \ .
\label{6}
\end{eqnarray} All four terms make contributions at all $p_T$,
although each is important in only restricted regions of $p_T$.
$ ({\cal SS})_1$ denotes two shower partons arising from one hard parton
(hence within one jet), and can be related through the use of $R$ to
the usual fragmentation that is described by the $D$ function.
$ ({\cal SS})_2$ denotes two shower partons that are from two separate but
nearby hard partons, and are therefore associated with two overlapping
jets.    $({\cal SS})_2$ is not expected to be important unless the
density of hard
partons is extremely high, such as that possibly at LHC. For brevity,
we shall write
$ ({\cal SS})_1$ simply as $\cal SS$, when no confusion is likely to arise.
  $\cal TT$ signifies two thermal partons whose
recombination yields the thermal hadrons, usually referred to as the
soft component.  $\cal TS$ denotes thermal-shower pairing and is the
new component that has never been considered before.  It turns out to
be important in the $3 < p_T < 8$ GeV range.  We emphasize that the
$\cal TS$ term would be absent if we do not treat the fragmentation of
a hard parton as the recombination of shower partons as done in
\cite{hy2}.  It is now evident that by considering $\cal S$ as the
showering effect of hard partons all hadrons are produced by
recombination, as is made explicit by Eqs.\ (\ref{5}) and (\ref{6}) in
the case of pions. Let us now specify what $\cal T$ and $\cal S$
represent.

\subsubsection{Thermal component}
   $\cal T$ is the thermal component including hydrodynamical flow.  It
is not our intention to derive soft parton distribution from
hydrodynamics.  We shall simply assume what is necessary to give rise
to the observed distribution of pions for $p_T < 2$ GeV. Since the
observed $dN_{\pi}/pdp$ at low $p$ is exponential, the necessary
invariant parton distribution for the thermal component is
\begin{eqnarray} {\cal T}(p_1) = p_1{dN_q^{\rm th}\over dp_1} =
Cp_1\exp (-p_1/T),
\label{7}
\end{eqnarray} where $T$ is the inverse slope enhanced by flow.
$C$ is the normalization factor to be adjusted to fit the pion data at
low $p$, and has the dimension $[{\rm momentum}]^{-1}$.  For
noncentral collisions,
which we do not consider in this paper, $C$ would depend on
centrality. We assume
that the   thermal partons are uncorrelated, except in special circumstances,
  so that we may write $F_{q\bar q'}^{\rm th}$ in the factorizable form
\begin{eqnarray}
   F_{q\bar q'}^{\rm th} (p_1, p_2) = {\cal T}(p_1) {\cal T}(p_2) = C^2
p_1 p_2 \exp [-(p_1+p_2)/T] \ .
\label{8}
\end{eqnarray} Substituting this into Eq.\ (\ref{5}) yields for the
thermal component of the pion distribution
\begin{eqnarray} {dN^{\rm th}_{\pi}  \over  pdp} =  {C^2 \over 6}
\exp (-p/T) ,
\label{9}
\end{eqnarray} which is the exponential form aimed for.  To fit the
data the parameters are
\begin{eqnarray} C = 23.2\ {\rm GeV}^{-1}, \qquad\qquad T = 0.317
\ {\rm GeV} ,
\label{10}
\end{eqnarray} as we shall see below.

\subsubsection{Shower partons}

Next we consider the shower distribution.  In a heavy-ion collision
let the parton $i$ be scattered into the transverse plane at rapidity
$y = 0$ with probability $f_i(k)$, i.e.,
\begin{eqnarray}
\left.{ dN^{\rm hard}_i \over  kdkdy}\right|_{y=0} = f_i (k)
\label{11}
\end{eqnarray} where $k$ is the transverse momenta of the parton. We
shall use the parametrization of $f_i (k)$ given in Ref.
\cite{sr}, obtained for the study of dilepton production in central
collision of gold nuclei at $\sqrt{s_{NN}} = 200$ GeV.  Due to energy
loss of the partons in the dense medium, not all partons emerge from
the reaction zone to hadronize outside.  Only a fraction of them do,
and we shall use $\xi$ to represent the effective fraction after
averaging over all central events such that $\xi f_i (k)$ denotes the
number of unquenched partons with momentum $k$ that are to hadronize.
Note that we do not refer to them as jets, since the notion of jets
presupposes that a hard parton fragments into a jet of hadrons.  That
supposition is, of course, just what we want to avoid.  We regard
$\xi$ as an effective fraction because we do not consider its
dependence on $k$ and do not delve into the space-time properties of
the hard scattering and subsequent evolution.  The value of $\xi$ will
be determined phenomenologically, and be regarded as an empirical
quantification of the degree of energy loss.

The hard parton $i$ at momentum $k$ creates a parton shower. Since
there are various types of shower partons for each type of initiating
parton $i$, let us use $S^j_i$ to denote the matrix of SPD's for $i
\to j$.  Although $i$ can be $u$, $d$, $s$, $\bar{u}$,
$\bar{d}$, $\bar{s}$ and $g$, $j$ is allowed to be quark and
antiquarks, but not gluon.  The role of gluons in the recombination
process has already been discussed earlier at the end of Sec.\ II-A. The
sea partons in the shower are saturated by gluon conversion. In the
notation of \cite{hy2} we use $K_{NS}$ to denote the valence quark in
the shower, $L \ (L_s)$ the light (strange) sea quarks in a
quark-initiated shower, and $G \ (G_s)$ the light (strange) sea quarks
in a gluon-initiated shower.  Thus the shower matrix $S^j_i$ has the
form
\begin{equation} S_i^j=\pmatrix{K&L&L_s\cr
                   L&K&L_s\cr
                 L&L&K_s\cr
                 G&G&G_s\cr} \ ,  \hspace{0.3cm} i=u,d,s,g,
\hspace{0.2cm}   j=u,d,s,     \label{10a}
\end{equation}
   where $K = K_{NS} + L$ and $K_s = K_{NS} + L_s$.  The anti-quarks
$\bar{u}$, $\bar{d}$ and $\bar{s}$ have the same structure, and are
related to $u$, $d$, $s$ as sea, and vice-versa. The parametrizations
for these SPD's have been completely determined in \cite{hy2} as
functions of the momentum fraction
$z$ of parton $j$ in parton $i$.
 Note that in Eq.\ (\ref{10a}) there
is no fourth column corresponding to gluons in the shower. The
determination of the five essential SPD's is carried out on the
condition of no shower gluons so that all hadrons specified by the
fragmentation functions are formed by  $q\bar q'$ recombination
without gluons. The underlying physics for this has already been discussed
in the last two paragraphs of Sec.\ II.A.

The distribution of shower parton $j$ with transverse momentum
$p_1$ in central heavy-ion collisions is then
\begin{eqnarray} {\cal S}(p_1) = \xi \sum_i \int^{\infty}_{k_0}dk k
f_i(k) S^j_i (p_1/k) \ ,
\label{11a}
\end{eqnarray} where $p_1$ and $k$ are collinear.  Since the input on
hard parton distribution $f_i(k)$ cannot be valid at low $k$, we shall
consider the above integral only for $k> k_0$, for which we set the
minimum at $k_0 = 3$ GeV.  In practice we cut off the upper limit of
integration at 20 GeV.

\subsubsection {Thermal-shower recombination}

With the shower partons specified we can now combine them with the
thermal partons to describe the $\cal TS$ term in Eq.\ (\ref{6}). We
have
\begin{equation} {\cal T}(p_1){\cal S}(p_2)=\xi\, C\, p_1 e^{-p_1/T}\
\sum_i\int dk\, k f_i(k)\,S_i^j(p_2/k) \ ,    \label{11b}
\end{equation}
where the distributions of thermal light quarks are
assumed to be flavor independent and the appropriate one is implied to
pair off with
$j$ to form the meson under consideration. The contribution to the pion
spectrum from thermal-shower recombination is then, using Eq.\
(\ref{5}),
\begin{equation} {dN_{\pi}^{\cal TS}\over pdp} = {1\over p^3} \int_0^p
dp_1 {\cal T}(p_1)\,{\cal S}(p-p_1) \ ,   \label{11c}
\end{equation} where a sum over $j$ is implied to match the flavors of
the valence quarks of the detected pion. Only the overall normalization
of this term depends on
$\xi$. The
$p$ dependence is our prediction, which is used to fit the data and
thereby determine $\xi$.

\subsubsection{Shower-shower recombination}

For two shower partons in the same jet we have
\begin{eqnarray} ({\cal SS})_1(p_1, p_2)=\xi\sum_i
\int dk k f_i(k)
\left\{S^j_i\left({p_1\over k}\right),S^{j'}_i\left({p_2\over
k-p_1}\right)
\right\},
\label{12}
\end{eqnarray} where the curly brackets signify the symmetrization of
the leading parton momentum fraction
\begin{widetext}
\begin{eqnarray}
\left\{S^j_i (z_1),\ S^{j'}_i \left({z_2\over 1-z_1}\right)\right\} =
{1\over  2} \left[S^j_i (z_1) S^{j'}_i \left({z_2\over 1-z_1}\right) +
S^j_i
\left({z_1\over 1-z_2}\right)S^{j'}_i (z_2)\right] .
\label{13}
\end{eqnarray}
We have shown in \cite{hy2} that the SPD's can be
determined from the recombination formula for the fragmentation
function
\begin{eqnarray} xD^M_i(x) = \int {dx_1  \over  x_1} {dx_2  \over x_2}
\left\{S^j_i (x_1),\ S^{j'}_i \left({x_2 \over 1-x_1}\right)
\right\} R_M (x_1, x_2, x) \ .
\label{14}
\end{eqnarray}
\end{widetext}
A sum over $j$ and $j'$ is implied in consort with the
$j$ and
$j'$ labels hidden in the RF that are relevant for $M$.  The
substitution of Eq.\ (\ref{12}) for the
$({\cal SS})_1$ term in Eq.\ (\ref{6}) into Eq.\ (\ref{1}) clearly yields
\begin{eqnarray} p{dN_M ^{\rm frag} \over  dp} = \xi \sum_i \int dk k
f_i(k) {p
\over  k} D^M_i\left( {p  \over  k}\right),
\label{15}
\end{eqnarray} which is the usual formula for the production of a
meson at high
$p_T$ in the fragmentation model, except for the presence of
$\xi$ here for reasons that have already been discussed above.

There is finally the possibility of recombination of two shower
partons from two different but partially overlapping showers.  The
corresponding $({\cal SS})_2$ term in Eq.\ (\ref{6}) should then be
\begin{widetext}
\begin{eqnarray} ({\cal SS})_2(p_1,p_2) = \delta_{y\phi} \xi ^2
\sum_{i,i'}
\int dk dk' kk'f_i(k) f_{i'}(k')\ S^j_i\left({p_1  \over k}\right)
S^{j'}_{i'}\left( {p_2  \over  k'}\right) ,
\label{16}
\end{eqnarray}
\end{widetext}
where a multiplicative factor $\delta_{y\phi}$ is
included to reflect the probability of overlap in $y$ and $\phi$ of
the two showers in order for collinear recombination of the partons
$j$ and $j'$ to take place.  The value of $\delta_{y\phi}$ can be
estimated by studying the size of the jet cone, and is expected to be
small.  Thus this mode of recombination is not likely to be important
at RHIC.  However, at very high energy, such as at LHC, where $f_i(k)$
is orders of magnitude higher, the $({\cal SS})_2$ term may well become
significant.

\begin{figure}[tbph]
\includegraphics[width=0.45\textwidth]{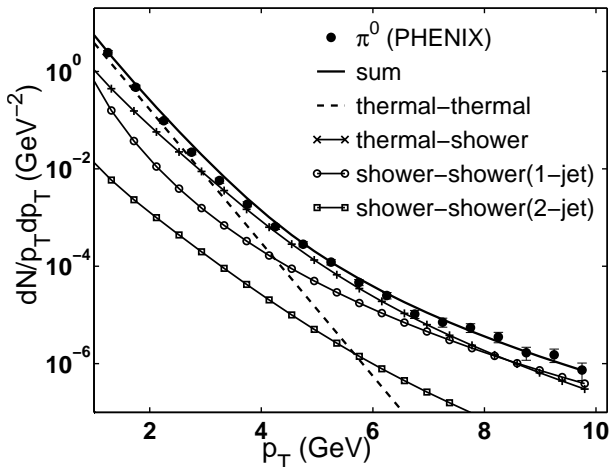}
\caption{Transverse momentum distribution of $\pi^0$ in Au-Au
collisions. Data are from \cite{ph}. The solid line is the sum of
four contributions to the recombination of partons: $\cal TT$ (dashed
line); $\cal TS$ (line with crosses); $({\cal SS})_1$, two shower
partons in one jet (line with open circles); $({\cal SS})_2$, two shower
partons from two overlapping jets (line with squares).}
\end{figure}

\subsubsection{Result on the pion spectrum}

Collecting all the pieces of Eq.\ (\ref{6}) together and substituting
them in Eq.\ (\ref{5}), we obtain the four contributions to the pion
spectrum.  The parameters $C$ and $T$ in Eq.\ (\ref{10}) are
determined by fitting the low-$p_T$ data. Ignoring the $({\cal SS})_2$
  contribution on the basis that
$\delta_{y\phi}$ is very small, there is only one parameter,
$\xi$, to adjust to fit the data for $p_T > 2$ GeV.  The result is
shown in Fig.\ 1.   With the value
\begin{eqnarray}
\xi = 0.07
\label{17}
\end{eqnarray} the fit of the $\pi ^0$ data from PHENIX
\cite{ph} on central Au-Au collisions at $\sqrt {s_{NN}}=200$ GeV is
excellent up to $p_T
\simeq 10$ GeV.  Note that in the region
$3 < p_T < 8$ GeV, the dominant contribution is from thermal-shower
recombination (line with crosses).  The conventional jet fragmentation is
from shower-shower recombination in one jet (line with circles); it
becomes more important than the
$\cal TS$ contribution only for $p_T > 9$ GeV.  To show the relative
size of the $({\cal SS})_2$ recombination from 2 jets, we assume
$\delta_{y\phi} = 0.01$ (a very rough estimate) just to put its
contribution on the figure.  It is indicated by the line with squares,
which is much lower than all others.

The result that thermal-shower dominates over shower-shower (1-jet)
recombination for $p_T<8$ GeV is our main finding in this work. It
shows the importance of considering the interaction between the
thermal partons and the partons created by hard scattering. That
interaction becomes particularly significant at the hadronization
scale where recombination occurs.

One way to see why $\cal TS$ is greater than $\cal SS$ in their
contributions to Eq.\ (\ref{1}) via (\ref{6}) is to examine their sum
in the following form
\begin{widetext}
\begin{eqnarray} {\cal TS}+{\cal SS}=\xi\,\sum_i\int dk\,k
f_i(k)\,S_i^{j'}\left({p_2\over k}\right)\
\left[Cp_1e^{-p_1/T}+S_i^j\left({p_1\over k-p_2}\right)\right]\ ,
\label{17a}
\end{eqnarray}
\end{widetext}
  putting aside the other symmetrizing term in ${\cal
SS}$ without any impediment to our argument. The first (thermal) term
inside the square bracket is much larger than the second (shower) term
when $p_1$ is small (but not infinitesimal),  which is a region of
$p_1$ that is relevant
for the soft parton to recombine with a shower parton at $p_2>3$ GeV
only if $R_\pi(p_1,p_2,p)$ is broad enough to encompass both. This
linking between soft and semi-hard partons not only enhances the
spectrum over simple fragmentation, but also modifies the structure of
what is usually regarded as minijet. Since there are no thermal soft
partons in $pp$ collisions, the pion distribution at intermediate and
high $p_T$ in $pp$ collisions must differ from that in $AA$ collisions.
Furthermore, the same-side correlations in the two cases are
necessarily different, as we shall discuss in Sec.\ 4.

\subsubsection{Dependence on the recombination function}
  To see how the
thermal-shower recombination depends on the broadness of RF, consider
the general form of the valon distribution
\begin{eqnarray} G(y_1,y_2)={1\over
B(a+1,a+1)}\,(y_1y_2)^a\,\delta(y_1+y_2-1).
\label{18}
\end{eqnarray}
Equations (\ref{3}) and (\ref{4}) follow from Eq.\ (\ref{18}) for $a=0$.
 If $a$ were larger, as one would expect for
the $\rho$ meson (since small $a$ corresponds to a tightly bound
state), then the valon distribution would be more sharply peaked at
$y_1=y_2=1/2$. In that case the recombination of a thermal parton at
low $p_1$ and a shower parton at intermediate $p_2$ is suppressed. To
make this point transparent, we show in Fig.\ 2(a) several possible
widths of a single-valon distribution $G(y)$ obtained from
$G(y_1,y_2)$ by  one integration, i.e.,
\begin{eqnarray} G(y)={1\over B(a+1,a+1)}\,[y(1-y)]^a,   \label{19}
\end{eqnarray}
which is normalized to 1 by one more integration. The
corresponding RF is given by Eqs.\ (\ref{4}) and (\ref{18}). When we
use that RF in Eq.\ (\ref{1}) and calculate  the distribution
$dN/pdp$ with only the $\cal TS$ contribution to $F_{q\bar q}$ taken
into account, the result is shown in Fig.\ 2(b) for
$a=0,1,2,5$. It is evident that the recombination of thermal-shower
partons is significantly suppressed at high $p_T$ when $a$ is large,
and becomes negligible when $R(p_1,p_2,p)$ tends toward being
proportional to $\delta(p_1-p/2)\delta(p_2-p/2)$.

\begin{figure}[tbph]
\includegraphics[width=0.45\textwidth]{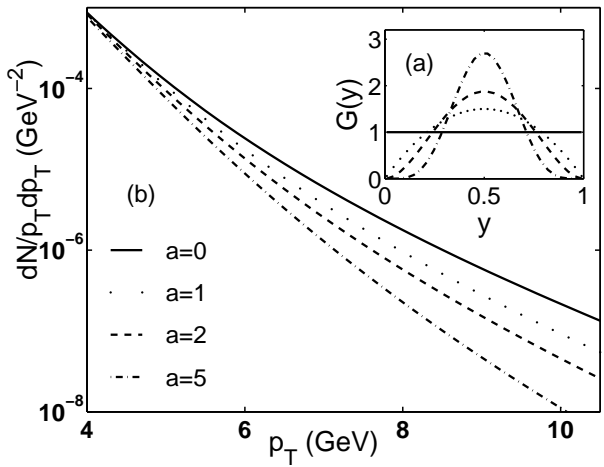}
\caption{(a) Various valon distributions in momentum fraction $y$
according to Eq.\ (\ref{19});\\
(b) The corresponding $p_T$ distributions from $\cal TS$
recombination.}
\end{figure}

\subsection{Energy loss}

In the absence of a space-time study of the problem that includes the
locations where hard collisions occur, it is not possible to consider
the medium effect on each and every hard parton that traverses the
medium. We have used the parameter $\xi$ to represent the overall
effect after averaging over all events in central collisions. The
value $\xi=0.07$ given in Eq.\ (\ref{17}) is a quantity deduced from
fitting the pion spectrum. It clearly indicates that not all hard
partons created in a heavy-ion collisions can get out of the dense
medium to hadronize. As suggested by the STAR data
\cite{star}, only those near the surface can escape the quenching
effect. Our result on the value of $\xi$ seems low by comparison to
the nuclear modification factor $R_{AA}(p_T)$, which is roughly 0.2
for $p_T>6$ GeV in central Au-Au collisions. To understand their
difference, let us examine what they measure respectively.

     $R_{AA}(p_T)$ is defined by the ratio
\begin{equation} R_{AA}(p_T)={dN/p_Tdp_T\,(AA)\over N_{coll}
dN/p_Tdp_T\,(pp)} \ ,    \label{21}
\end{equation} where $N_{coll}$ is the average number of binary
collisions. In the denominator the production of particles at high
$p_T$ in $pp$ collisions can be well described by the fragmentation of
hard partons. However, we have seen that the usual fragmentation
corresponds to the shower-shower recombination in 1-jet, which is not
the important part of hadronization in $AA$ collisions for $3<p_T<8$
GeV. The suppression factor $\xi$ may, in the spirit of Eq.\
(\ref{21}), be written in the schematic form
\begin{equation}
\xi=\left<{dN/p_Tdp_T\,(AA)\over \int{\cal T}\hat{\cal S}
R\,(AA)}\right>
\ ,
\label{22}
\end{equation} where $\hat{\cal S}$ is the shower component expressed
in Eq.\ (\ref{11a}), but without the $\xi$ factor. The angular
brackets denote an average over all \pt. The denominator is the
thermal-shower recombination in $AA$ collisions, if all hard partons
get out of the medium to  hadronize. It is then clear why
$\xi$ is smaller than $R_{AA}$. It  should not only account for the
suppressed number of hard partons in the numerator that get out from
the dense medium to hadronize (which $R_{AA}$ does also), but is also
made smaller by the denominator where the
${\cal T}\hat{\cal S}$ recombination  is larger than the scaled
contribution from $pp$ collisions.

\begin{figure}[tbph]
\includegraphics[width=0.45\textwidth]{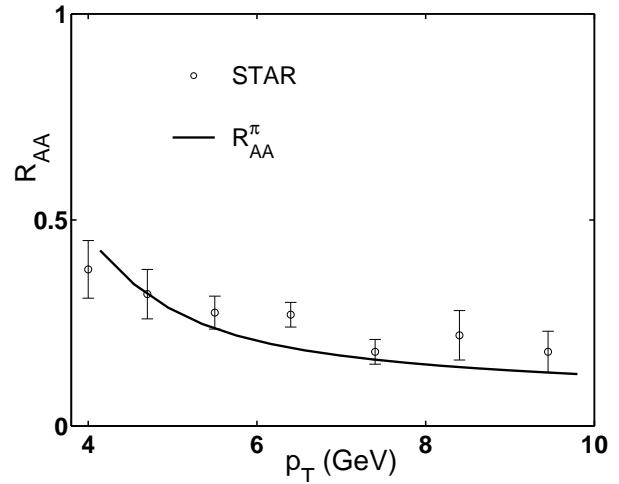}
\caption{Nuclear modification factor $R_{AA}$ compared to the
calculated result using  only pions, $R_{AA}^\pi$.}
\end{figure}

Another way to exhibit the effect of energy loss on our result is to
calculate $R_{AA}(p_T)$ directly from our pion distribution, assuming
that the pions dominate over all other particles. That assumption is
  invalid for $p_T<4$ GeV, since the production of proton is known
to be roughly equal to that of pion in the $3<p_T<4$ GeV region.
Nevertheless, it is illuminating to see what the pion component gives
(call it $R_{AA}^\pi$) relative to the experimental $R_{AA}$ for
$p_T>4$ GeV. Since ${\cal SS}$ recombination is just the
fragmentation component in $AA$ collisions as stated in Eq.\
(\ref{15}), we can obtain $N_{\rm coll}dN_\pi/p_Tdp_T(pp)$ simply by
omitting the factor $\xi$ from $dN_\pi^{\cal SS}/pdp$. That is, we
have
\begin{eqnarray}
R_{AA}^\pi(p) = {dN_\pi/pdp \over \xi^{-1}dN_\pi^{\cal SS}/pdp} \ ,
\label{22aa}
\end{eqnarray}
where the numerator corresponds to the solid line in Fig.\ 1 and
$dN_\pi^{\cal SS}/pdp$ corresponds to the line with open circles in
the same figure. The result for $R_{AA}^\pi$ is shown in Fig.\ 3 by
the solid line, which is in reasonable agreement with the data on
$R_{AA}$ \cite{st2} that is not limited to the pions. It is now clear
that the ratio of those two lines is equal to the ratio
$R_{AA}^\pi/\xi$, which in turn is approximately equal to
$R_{AA}/\xi$. Since the thermal-thermal recombination is negligible
for $p_T>4$ GeV, the ratio $(dN_\pi/pdp)/ (dN_\pi^{\cal
SS}/pdp)$ is essentially independent of $\xi$. Thus
$R_{AA}^\pi/\xi$ is roughly independent of energy loss and therefore
of centrality.

It is worth remarking  on how the nuclear-size effect enters into our
one-dimensional treatment of the hadronization process. We emphasize
that hadronization occurs outside the volume of dense matter and along
the direction of the detected particle. The 3D nuclear-size effect
influences the properties of the partons {\it before} they enter into
the interaction region for hadronization, and is therefore not explicitly
present in the recombination formula. The recombining partons are either
the soft thermal ones or the semi-hard shower partons. Centrality obviously
affects the magnitude of the thermal source, i.e., $C$, and the degree of
energy loss, i.e., $\xi$. We have made preliminary study of the centrality
dependence of $dN_{\pi}/pdp$ and found agreement with the data by appropriate
adjustment of $C$ and $\xi$ in essentially the same way as we have done in this
paper for the most central collisions. We mention this only to point out that
the 1D description of recombination does not preclude the possibility
of accounting
for the nuclear-size effect, which is manifestly 3D, but occurs prior to the
hadronization process.

The suppression factor $\xi$, as expressed in Eq.\ (\ref{22}), cannot
be measured directly. There is, however, a measurable quantity  that is
closely related to $\xi$. Since only the hard partons that are created
near the surface of the collision region can get out, most jets
detected in $AA$ collisions do not have a partner in the opposite
direction. That does not mean the non-existence of back-to-back jets.
A pair of hard partons that are created near the edge, but directed at
around
$90^{\circ}$ relative to the radial position of creation measured from
the center, and yet remain in the transverse plane (i.e., roughly
tangent to the cylinder), do not go through the bulk of the medium.
Those two hard partons can then lead to back-to-back jets that are
detectable. It is then of interest to measure the total number of
events that contain back-to-back jets. Let $R_{jj/j}(\bar p_T)$ be the
ratio of events with back-to-back jets to the total number of events
containing any jets, where $\bar p_T$ is the minimum value of \pt that
any particle must have to be counted as part of a jet. Thus, for
example, for $\bar p_T=5$ GeV, $R_{jj/j}(\bar p_T)$ refers to the
fraction of events  containing two particles with
$p_T>5$ GeV in opposite directions out of all events with any particle
having
$p_T>5$ GeV.  $R_{jj/j}(\bar p_T)$ is then a measure of the fraction of
space near the surface that can give rise to particles at high \pt
through thermal-shower recombination. It can have dependence on
centrality. The precise relationship between
$\xi$ and $R_{jj/j}$ is a separate problem worthy of detailed
investigation, especially if the experimental determination of
$R_{jj/j}$ is forthcoming.

In a Monte Carlo calculation, such as that employed in \cite{gr}, where
space-time trajectories of the hard partons are tracked, it is
possible to compute the parton momenta after energy losses are taken
into account. Those emergent partons then generate shower partons
whose recombination with thermal partons presumably results in a yield
that can check the value of our mean suppression factor $\xi$. In our
treatment in the momentum space only, we determine $\xi$ by fitting
the pion data, but once fixed the relative magnitude between $\cal TS$
and ${\cal SS}$ is also fixed. Moreover, there is no more freedom in
the determination of the inclusive distribution of other particles,
such as kaon and proton.

\subsection{Kaon production}

For the production of kaon the RF has the explicit form that follows
from Eq.\ (\ref{2}) with $a=1$ and $b=2$ \cite{hy3}
\begin{equation}
R_K(p_1,p_2,p)=12\,{p_1^2p_2^3\over p^4}\,\delta(p_1+p_2-p) \ .
\label{22a}
\end{equation}
Using this in Eq.\ (\ref{1}) yields for $K^+$
\begin{eqnarray}
{dN_{K^+}\over pdp}={12\over p^6}\,\int_0^p
dp_1\,p_1(p-p_1)^2\,F_{u\bar s}(p_1,p-p_1) \ .
\label{22b}
\end{eqnarray}
Note that because the kaon is not as tightly bound as pion, the RF is
not as broad as that for pion, with the consequence that the factor
$p_1(p-p_1)^2$ in the integrand forces the $u$ and $\bar s$ quarks to
have closer momenta than those in Eq.\ (\ref{1}).

There are four terms for $F_{u\bar s}$ as in Eq.\ (\ref{6}). The
calculational procedure is basically the same as for pion. A slight
complication arises from the strange quark being different from the
light quarks. In the thermal component we shall simply attach a
multiplicative factor $\lambda_s$ for the $s$ sector
\begin{eqnarray}
{\cal T}_s = \lambda_s\,{\cal T}\ ,  \label{22c}
\end{eqnarray}
where we set $\lambda_s$ to be the Wr{\` o}blewski factor at
$\lambda_s=0.5$ \cite{brs}. The thermal-shower recombination now has
two terms
\begin{widetext}
\begin{eqnarray}
{\cal T}(p_1){\cal S}_s(p_2)+{\cal S}(p_1){\cal T}_s(p_2)
  =\xi \,
C\,\sum_i\int dk\,kf_i(k)\left[p_1e^{-p_1/T}\,S_i^{\bar
s}\left({p_2\over
k}\right)+\lambda_sp_2e^{-p_2/T}S_i^u\left({p_1\over k}\right)
\right].       \label{22d}
\end{eqnarray}
\end{widetext}
The $({\cal SS})_1$ and $({\cal SS})_2$ contributions are as expressed in
Eqs.\ (\ref{12}) and (\ref{16}), respectively, with $\{j,j'\}$
identified as $\{u,\bar s\}$.

\begin{figure}[tbph]
\includegraphics[width=0.45\textwidth]{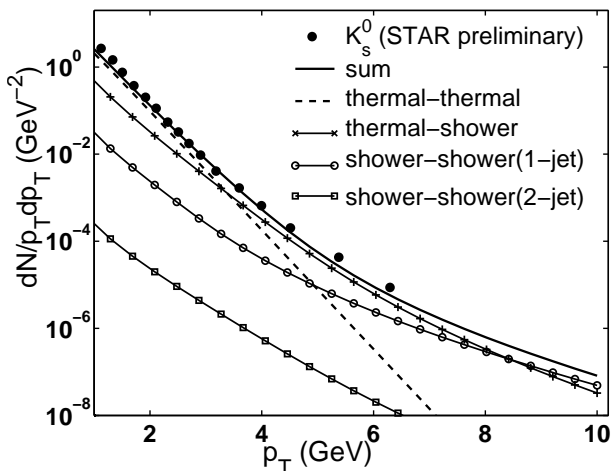}
\caption{Transverse momentum distribution of $K^+$ in Au-Au
collisions. Data for $K_s^0$ are preliminary from \cite{st3}. The
symbols for the four contributions are the same as in Fig.\ 1.}
\end{figure}

The four contributions to the kaon spectrum are shown in Fig.\ 4. No
adjustable parameter has been used beyond what has already been
determined in the previous section. The agreement between the sum
(solid line) and the data is evidently very good. The data are for
$K_s^0$ at 0-5\% centrality and extend to $p_T\sim 6$ GeV
\cite{st3}, farther than for $K^+$ \cite{ppi,star3}. As in the case of
pions the thermal-shower recombination is more important than both
thermal-thermal and shower-shower (1-jet) for $3<p_T<8$ GeV. As in
Fig.\ 1 the shower-shower (2-jet) curve is shown for
$\delta_{y\phi}=0.01$ and is negligible. There are some small
differences in the various components of the pion and kaon spectra,
but on the whole the two are basically similar.

\section {Baryon Production by Recombination}

\subsection{General considerations}

The extension of the treatment in the preceding section to baryon
production is conceptually straightforward. The generalization of
Eq.\ (\ref{1}) is clearly
\begin{eqnarray}
p{dN_B\over dp}=\int {dp_1\over p_1}{dp_2\over p_2}{dp_3\over p_3}\,
F(p_1,p_2,p_3)\,R_B(p_1,p_2,p_3,p)\ ,
\label{23}
\end{eqnarray}
where $F(p_1,p_2,p_3)$ is the joint distribution of three relevant
quarks to form the baryon $B$. The RF is related to the non-invariant
valon distribution by
\begin{eqnarray}
R_B(p_1,p_2,p_3,p)=g_{st}\,y_1y_2y_3\,G_B(y_1,y_2,y_3),
y_i=p_i/p,
\label{24}
\end{eqnarray}
where $g_{st}$ is a statistical factor, and $G_B(y_1,y_2,y_3)$ has the
general form
\begin{eqnarray}
G_B(y_1,y_2,y_3)=g_B\,y_1^\alpha y_2^\beta y_3^\gamma\,
\delta(y_1+y_2+y_3-1)\ ,
\label{25}  \\
g_B=[B(\alpha+1,\beta+\gamma+2)\,B(\beta+1,\gamma+1)]^{-1}\ .
\label{26}
\end{eqnarray}

For proton, $y_1$ and $y_2$ refer to the momentum fractions of the
$U$ valons, and $y_3$ to that of the $D$ valon. The exponents
$\alpha, \beta,$ and $\gamma$ have been determined in \cite{hy4} to be
\begin{eqnarray}
\alpha=\beta=1.75, \hspace{1cm} \gamma=1.05
\label{27}
\end{eqnarray}
from parton distribution functions of the proton. Note that although
the $U$ and $D$ valons have the same constituent quark masses, they
have different average momentum fractions calculable from Eq.\
(\ref{25}) \cite{hy4}
\begin{eqnarray}
\left<y\right>_U=0.3644, \hspace{1cm} \left<y\right>_D=0.2712\ ,
\label{28}
\end{eqnarray}
which satisfies the sum rule
\begin{eqnarray}
2\,\left<y\right>_U+\left<y\right>_D=1\ .
\label{29}
\end{eqnarray}
For hyperons, the lack of information about their parton distribution
functions deprives us of any such detail knowledge of their valon
distributions, and consequently of their RF's. Nevertheless, on the
basis  that the constituent quarks in all baryons are not tightly
bound, we expect the exponents $\alpha, \beta$ and $\gamma$ for
hyperons to be also in the range between 1 and 2.

The 3-quark distribution now has more terms in the various possible
contributions from the thermal and shower partons. Schematically, it
takes the form
\begin{eqnarray}
F_{qq'q''} &=& {\cal TTT+TTS+T}({\cal SS})_1+({\cal SSS})_1\nonumber\\
&&+{\cal T}({\cal SS})_2+({\cal
S}({\cal SS})_1)_2+({\cal SSS})_3\ .
\label{30}
\end{eqnarray}
They are arranged in increasing order of the number of hard partons
involved: the first term has none, the next three one, the following
two two, and the last term three.  We shall consider only the first four
terms, since they involve thermal partons and the shower of only one
hard parton. The fifth term involves partons from two overlapping jets,
and is ignored at RHIC energy despite the enhancement by $\cal T$.

\subsection{Proton production}

Let us focus our attention on proton production at high $p_T$. As
before, we omit the subscript $T$ in referring to momenta in the
transverse plane, and obtain from Eqs.\ (\ref{23})-(\ref{26})
\begin{widetext}
\begin{eqnarray}
{dN_p\over pdp}={g_p\,g_{st}\over p^{2\alpha+\gamma+4}} \int_0^p\,dp_1
\int_0^{p-p_1}
dp_2\,(p_1p_2)^\alpha(p-p_1-p_2)^\gamma\,F(p_1,p_2,p-p_1-p_2)\ .
\label{30a}
\end{eqnarray}
\end{widetext}
The ${\cal TTT}$ contribution can be computed analytically. Using Eq.\
(\ref{7}) for each ${\cal T}$, we obtain for the thermal spectrum
\begin{eqnarray}
{dN_p^{\rm th}\over pdp}={C^3\over 6}
p\,e^{-p/T}\,
{B(\alpha+2,\gamma+2)B(\alpha+2,\alpha+\gamma+4)\over
B(\alpha+1,\gamma+1)B(\alpha+1,\alpha+\gamma+2)}\ ,
\label{31}
\end{eqnarray}
where $C$ and $T$ are given in Eq.\ (\ref{10}), and $\alpha, \gamma$
in (\ref{27}). The statistical factor $g_{st}$ is 1/6 when the
spin-flavor consideration is taken into account \cite{hy}. Equation
(\ref{31}) is not reliable at very small $p$, since the proton mass
effect invalidates our essentially scale-invariant formulation for
relativistic particles. Furthermore, at small $p_T$ the problem becomes
3D, so our treatment should be modified accordingly.

The $\cal TTS$ contribution is
\begin{widetext}
\begin{eqnarray}
{dN_p^{\cal TTS}\over pdp}={g_p\,C^2\,\xi\over
6p^{2\alpha+\gamma+4}}\,\int
dp_1dp_2\,(p_1p_2)^{\alpha+1}(p-p_1-p_2)^{\gamma+1}\sum_i\int dk\,k\,
f_i(k)\,U_i(k,p_1,p_2,p)\ ,
\label{32}
\end{eqnarray}
where
\begin{eqnarray}
U_i(k,p_1,p_2,p)={1\over p_1}\,e^{-(p-p_1)/T}\,S_i^u\left({p_1\over
k}\right)+{1\over p_2}\,e^{-(p-p_2)/T}\,S_i^u\left({p_2\over
k}\right)
+ {1\over
p-p_1-p_2}\,e^{-(p_1+p_2)/T}\,S_i^d\left({p-p_1-p_2\over k}\right)\ .
\label{33}
\end{eqnarray}
The $\cal TSS$ contribution is
\begin{eqnarray}
{dN_p^{\cal TSS}\over pdp}={g_p\,C\,\xi\over
6p^{2\alpha+\gamma+4}}\,\int
dp_1dp_2\,(p_1p_2)^{\alpha}(p-p_1-p_2)^{\gamma}\sum_i\int dk\,k\,
f_i(k)\,V_i(k,p_1,p_2,p)\ ,
\label{34}
\end{eqnarray}
where
\begin{eqnarray}
V_i(k,p_1,p_2,p)&=&p_1\,e^{-p_1/T}\,\left\{S_i^u\left({p_2\over
k}\right),\ S_i^d\left({p-p_1-p_2\over k-p_2}\right)\right\}
+p_2\,e^{-p_2/T}\,\left\{S_i^u\left({p_1\over
k}\right),\ S_i^d\left({p-p_1-p_2\over k-p_1}\right)\right\}
\nonumber\\
  &+&
(p-p_1-p_2)\,e^{-(p-p_1-p_2)/T}\,\left\{S_i^u\left({p_1\over
k}\right),\ S_i^u\left({p_2\over k-p_1}\right)\right\} \ .
\label{35}
\end{eqnarray}
\end{widetext}
The curly brackets in Eq.\ (\ref{35}) denote symmetrization as in
Eq.\ (\ref{13}). Finally, we also have ${\cal SSS}$ which is simply
related to the fragmentation of a hard parton into proton. The latter
has already been studied in \cite{hy2}. The corresponding FF,
$D_i^p(z)$, can be used here as in Eq.\ (\ref{15}) to give
\begin{eqnarray}
{dN_p^{\cal SSS}\over pdp}={\xi\over p}\sum_i\int dk\,
f_i(k)\,D_i^p\left({p\over k}\right)\ .
\label{37}
\end{eqnarray}

\begin{figure}[tbph]
\includegraphics[width=0.45\textwidth]{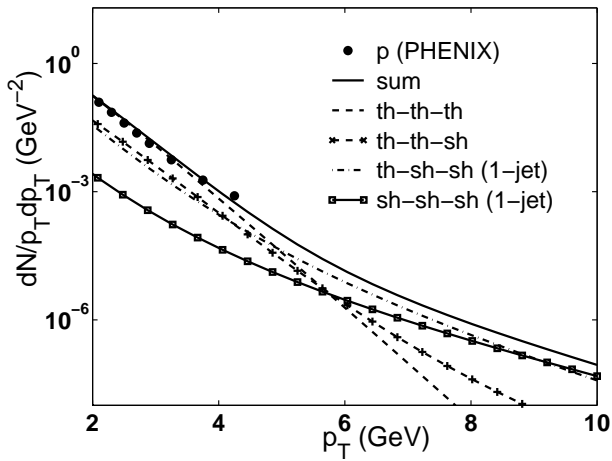}
\caption{Transverse momentum distribution of proton in Au-Au
collisions. Data are from \cite{ph}. The solid line is the sum of
four contributions to the recombination of partons: $\cal TTT$ (dashed
line); $\cal TTS$ (line with crosses); ${\cal T}{\cal SS}$, one
thermal parton with two shower partons in one jet (dashed-dot line);
${\cal SSS}$, three shower partons from one jet (line with
squares).}
\end{figure}

The result of our calculation for the four types of contributions are
shown separately in Fig.\ 5. Their sum is shown as solid line, and
agrees well with the data from PHENIX in Au-Au collisions
at $\sqrt {s_{NN}}=200$ GeV for 0-5\% centrality \cite{ppi}. We
emphasize that there are no free parameters to adjust to achieve the
good agreement.  The result below
$p_T=2$ GeV is not compared with data, since the proton mass effect
becomes important there and invalidates our scale-invariant
formulation.  All three contributions that involve thermal partons
have about the same strength at
$p_T\approx 4$ GeV. What dominates at higher \pt is the
${\cal T}{\cal SS}$ component until $p_T > 9$ GeV where the ${\cal
SSS}$ component takes over. The data available now are only for \pt
up to 4.2 GeV and therefore cannot check our prediction in detail. At
this point we can only conclude that the departure from the
exponential behavior in the data for $p_T>3.5$ GeV is well accounted
for by the two types of thermal-shower recombination, but not by
direct hard-parton fragmentation. That may be regarded as empirical
support for the role of thermal-shower recombination.

Having obtained both the proton spectrum here and the pion spectrum
in Sec.\ 2, we can now calculate the $p/\pi$ ratio and compare it to
the data. Since the calculated result shown in Fig.\ 5 is only for
$p_T\ge2$ GeV/c due to our neglect of the proton mass $m_p$ in our
formulation, the $p/\pi$ ratio that is represented by the solid line in
Fig.\ 6 is not shown for $p_T<2$ GeV/c.  It has the broad feature that
the ratio is greater than 1 at $p_T\approx 3$ GeV/c.
For low $p_T$  where the proton mass effect is important we adopt the
Ansatz by replacing the factor $p^{-(2\alpha+\gamma)}$ in Eq.\
(\ref{30a}), which arises from $G_B$ in Eq.\ (\ref{25}), by
$m_T^{-(2\alpha+\gamma)}$, where $m_T=(m_p^2+p^2)^{1/2}$. The result
is shown as dashed line in Fig.\ 6
and exhibits excellent agreement with the data  \cite{ppi,adl}.
It should be noted that the amendment is at best kinematical, since no dynamical
effect at low $p_T$ has been considered.
What is  noteworthy is that the maximum exceeds 1 at the peak, which
is the anomaly that the fragmentation model cannot explain. The slow
decrease of the ratio as
\pt increases is a prediction of our model that can be checked
by future data at $p_T>3$ GeV/c.

\begin{figure}[tbph]
\includegraphics[width=0.45\textwidth]{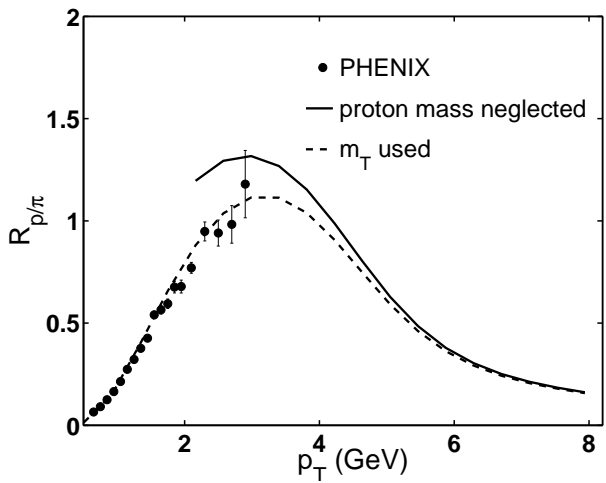}
\caption{ Comparison of calculated $p/\pi$ ratio with data from
\cite{ppi}. The solid line is the ratio of the solid lines in Figs.\ 5 and 1,
without the mass effect taken into account. The dashed line is the result after
$p$ is replaced by $m_T$ in the way described in the text.}
\end{figure}

\section{Same-side correlation}

We have stated that because of the thermal-shower recombination the
structure of jets in $AA$ collisions must be different from that in
$pp$ collisions. One way to probe that difference is to study
same-side correlation of particles. In \cite{hy2} we have calculated
the 2-pion correlated distribution in a $u$-quark initiated jet. That
type of calculation can be incorporated into the study of same-side
correlation in $pp$ collisions by integrating over hard-parton
momentum and summing over all types of hard partons. For $AA$
collisions one would consider $2q$ and 2$\bar q$ from thermal and
shower partons and then recombine them to form two pions. Quantitative
computation of the process will not be attempted here. Instead, we
make some qualitative remarks in light of some data from RHIC that
has some  relevance.

In \cite{star} a quantity $I_{AA}$ is defined to be the ratio of all
charged particles (above a modeled background) within a range of
azimuthal angle $\Delta\phi$ around the trigger particle in Au-Au
collisions to the same quantity (without background) in $pp$
collisions. If there is no medium effect and if jets in $AA$ and $pp$
collisions are the same, then $I_{AA}$ should be 1. It is found in
\cite{star} that if the trigger momentum is in the range $4<p_T^{\rm
trig}<6$ GeV, the value of \iaa \ in central collisions is consistent
with 1, though being more like 1.1. However, for $3<p_T^{\rm
trig}<4$ GeV, \iaa is significantly higher, roughly 1.5 for $N_{\rm
part}$ from 150 to 350. The former case for higher $p_T^{\rm trig}$
has been regarded as evidence that the nearside azimuthal peaks in
$AA$ and $pp$ collisions are similar
\cite{star,star2}, but no explanation is given for the discrepancy
when $p_T^{\rm trig}$ is lower. The azimuthal distribution of the
nearside particles is not shown in \cite{star,star2} for $3<p_T^{\rm
trig}<4$ GeV. It is also not clear for that case whether the
particles associated with the trigger are integrated over the range
$2<p_T<p_T^{\rm trig}$ only, as it is stated explicitly for the case
of \46 GeV \cite{star2}.

In our view there is no reason why \iaa should be equal to 1 for the
same-side particles. Indeed, we regard the value $I_{AA}\approx 1.5$
for $3<p_T^{\rm trig}<4$ GeV to be an  evidence in support of
the enhancement due to thermal-shower recombination. The question
that we should address is why \iaa is lower at higher $p_T^{\rm
trig}$.

Although our formalism does not facilitate the calculation of the
azimuthal distribution, we can regard the integral over $\Delta\phi$
under the same-side peak that enters the determination of \iaa as
being roughly equal to an appropriate integral of the two-particle
distribution $dN/dP_1dP_2$ over $P_2$, with $P_1$ being assigned the
role of the trigger momentum. Despite our restriction to collinear
momenta, the integral $\int dP_2$ accounts for all associated
particles, as does $\int d\Delta\phi$ in the data analysis by STAR
\cite{star}. In our approach we have a hard parton at $k$ creating a
shower in which two partons at $p_1$ and $p_2$ combine separately with
two thermal partons at $p_1'$ and $p_2'$ to form two pions at
$P_1=p_1+p_1'$ and $P_2=p_2+p_2'$. The distribution is
\begin{widetext}
\begin{eqnarray}
P_1P_2{dN_{\pi\pi}\over dP_1dP_2}=\int {dp_1\over p_1}{dp_2\over
p_2}{dp_1'\over p_1'}{dp_2'\over p_2'} ({\cal SS})_1(p_1,p_2) {\cal
T}(p_1'){\cal T}(p_2')R_\pi(p_1,p_1',P_1)R_\pi(p_2,p_2',P_2) \ .
\label{37a}
\end{eqnarray}
\end{widetext}
For $pp$ collisions the thermal partons are replaced by other shower
partons initiated by the same hard parton that
produces the ones at $p_1$ and $p_2$, i.e., in Eq.\ (\ref{37a}) we
make the replacement
\begin{eqnarray}
({\cal SS})_1(p_1,p_2){\cal T}(p_1'){\cal T}(p_2') \rightarrow ({\cal
SSSS})_1(p_1,p_2,p_1',p_2')/\xi \ ,
\label{38}
\end{eqnarray}
where $({\cal SSSS})_1$ is a generalization of $({\cal SS})_1$ in
Eq.\ (\ref{12}) to 4 shower partons.

We now see that in $pp$ collisions when  the trigger momentum, $P_1$,
is high, it forces $p_1$ or $p_1'$  in
$({\cal SSSS})_1(p_1,p_2,p_1',p_2')$ to be high, with the consequence
that $p_2$ and $p_2'$ must be low.  Shower partons $S_i^j(z)$ at small
momentum fraction $z$ are known to have high density. They are shown
in Fig.\ 2 of Ref.\ \cite{hy2}. Indeed, the 2-pion distribution in a
hard parton based on $({\cal SSSS})_1$ has been calculated in \cite{hy2},
where it is shown that the distribution for the momentum fraction
$X_2=P_2/k$ of the (non-trigger) second particle becomes very high at
low $X_2$. The high density at low $p_2'$ renders $({\cal SSSS})_1$ to be
of the same order as ${\cal T}(p_2')$ in $AA$ collision, resulting in
an increase of the $\pi\pi$ distribution in $pp$ collisions. For that
reason \iaa is lower (near 1) at higher $p_T^{\rm trig}$. In other
words, since \iaa is dominated by the large number of particles in the
low \pt range in the jets, it masks the differences in the structures
of the jets (at intermediate and high $p_T$) produced in nuclear and
hadronic collisions.

The drawback in using \iaa as a measure of same-side correlation is
that it involves integrations in both  $\Delta\phi$ and $P_2$, a
procedure that is likely to suppress the distinctive features of
thermal-shower recombination. It is recommended that the momenta of
all particles in a jet are projected along the trigger momentum,
$P_1$, and then the distribution in those projected momenta, $P_2$, is
determined for several values of  $P_1$.

\section{Conclusion}

By showing the importance of considering shower partons created by
hard partons, we have called into question the conventional paradigm
in particle production at high \pt in heavy-ion collisions. The usual
approach is to regard such particles as the products of parton
fragmentation. We have shown that all particles are the result of
parton recombination, including but not limited to  the ones usually
regarded as fragments. The important input that makes feasible our
approach based entirely on recombination is the shower parton
distributions derived in Ref.\ \cite{hy2}. Those distributions are
determined by analyzing the fragmentation functions for
parton-initiated jets in the recombination model. Once they are known,
it is conceptually unavoidable to consider the recombination of
thermal and shower partons in heavy-ion collisions. The result of such
subprocesses turns out to be very important in the $3<p_T<8$ GeV
range, as we have shown. The usual subprocesses of hard scattering
followed by fragmentation are found to be unimportant until higher
$p_T$. That results in a paradigm shift that has far reaching
consequences.

The phenomenon of energy loss of partons traversing dense medium can
be related to experimental observables only by means of some valid
model of hadronization that connects the partons to hadrons. If
fragmentation is not important in the region of \pt under
investigation, then serious modification of the quantitative
implications of jet quenching must be considered, since the content of
a jet has been altered from that produced in $pp$ collisions. What
such a modification should be has not been studied in this paper. We
have only used an effective parameter
$\xi$ to account for the fact that not all hard partons can get out
of the dense medium to hadronize.  We admit that such a method of
treating energy loss is very rudimentary in this first attempt to
study the effects of shower partons. Yet we have found consistency in
being able to reproduce the spectra of pion, kaon, and proton. A more
detailed investigation that tracks the space-time history of the
produced partons would require an evolution code that is
beyond the scope of this paper.

Particles that are formed by thermal-shower recombination are part of
a jet produced in heavy-ion collisions but are not present in jets in
$pp$ collisions. Thus the structures of jets produced in $AA$ and
$pp$ collisions are different. To find evidences for such differences
is of paramount importance in both experimental and theoretical work
to follow. The ratio \iaa discussed in Sec. 4 on same-side correlation
reveals a limited glimpse of that difference. We have suggested
that the measurement of 2-particle inclusive distribution at high \pt
would provide a more accurate description of the jet structure and
can be used to check the prediction that can be made in our framework.

At this point the single-particle spectra that we have considered
provide sufficient encouragement from the agreement with existing
data to suggest that the recombination approach has captured the
essence of hadronization at any \pt and that shower partons play an
important role in the process. More data at higher
\pt and on other species will give more stringent tests that the
recombination model must pass in order to establish the solidity
required for a reliable mechanism of hadronization.

\section*{Acknowledgment}
We thank the participants of the Mini-Workshop on Quark
Recombination for their helpful comments, especially V.\ Greco,
C.\ M.\ Ko and B.\ Mueller.  This work was supported, in part,  by the
U.\ S.\ Department of Energy under Grant No. DE-FG03-96ER40972  and by
the Ministry of Education of China under Grant No. 03113.



\begin{thebibliography}{99}

\bibitem{hy} R.\ C.\ Hwa, and C.\ B.\ Yang, Phys.\ Rev.\ C {\bf 67},
034902 (2003).

\bibitem{gr} V.\ Greco, C.\ M.\ Ko, and P.\ L\'{e}vai, Phys.\ Rev.\
Lett.\ {\bf 90}, 202302 (2003); Phys.\ Rev.\ C {\bf 68}, 034904 (2003).

\bibitem{fr} R.\ J.\ Fries, B. M\"{u}ller, C.\ Nonaka and S.\ A.\
Bass, Phys.\ Rev.\ Lett.\ {\bf 90}, 202303 (2003);  Phys.\ Rev.\ C
{\bf 68}, 044902 (2003).

\bibitem{ppi} S.\ S.\ Adler, PHENIX Collaboration, Phys.\ Rev.\ C
{\bf 69}, 034909 (2004).

\bibitem{v2} D.\ Moln{\' a}r and S.\ A.\ Voloshin,  Phys.\ Rev.\
Lett.\ {\bf 91}, 092301 (2003).

\bibitem{sor}P.\ Sorensen, STAR Collaboration, nucl-ex/0305008.

\bibitem{hy2}
    R.\ C.\ Hwa and C.\ B.\ Yang, hep-ph/0312271.

\bibitem{ht} R.\ C. Hwa, Nucl.\ Phys.\ B (Proc.\ Suppl.) {\bf 92},
348 (2001).

\bibitem{wa} X.\ N.\ Wang and M.\ Gyulassy, Phys.\ Rev.\ Lett.\ {\bf
68}, 1480 (1992).

\bibitem{gy} For a recent review see M.\ Gyulassy, I.\ Vitev, X.\ N.\
Wang, and B.\ W.\ Zhang, in {\it Quark Gluon Plasma 3}, edited by R.\
C. Hwa and X.\ N.\ Wang (World Scientific, Singapore, 2004).

\bibitem{jq} X.\ N.\ Wang, Phys.\ Rev.\ C {\bf 58}, 2321 (1998);
{\bf 61}, 064910 (2000).

\bibitem{hw} R.\ C.\ Hwa, Phys.\ Rev.\ D {\bf 22}, 1593 (1980).

\bibitem{hy3} R.\ C.\ Hwa and C.\ B.\ Yang, Phys.\ Rev.\ C {\bf 66},
025205 (2002).

\bibitem{su}P.\ J.\ Sutton, A.\ D.\ Martin, R.\ G.\ Roberts,
and W.\ J.\ Stirling, Phys.\ Rev.\ D {\bf 45}, 2349 (1992).
\bibitem{sr} D.\ K.\ Srivastava, C.\ Gale, and R.\ J.\ Fries, Phys.\
Rev.\ C {\bf 67}, 034903 (2003).

\bibitem{ph} S.\ S.\ Adler, PHENIX Collaboration, Phys.\ Rev.\
Lett.\  {\bf 91}, 072301 (2003).

\bibitem{star} C.\ Adler {\it et al.}, STAR Collaboration,  Phys.\
Rev.\ Lett.\  {\bf 90}, 082302 (2003).

\bibitem{st2} J.\ Adams {\it et al.}, STAR Collaboration, Phys.\ Rev.\
Lett.\  {\bf 91}, 172302 (2003).

\bibitem{brs} P.\ Braun-Munzinger, K.\ Redlich, and J.\ Stachel, in
{\it Quark Gluon Plasma 3}, edited by R.\ C.\ Hwa and X.\ N.\ Wang,
(World Scientific, Singapore, 2004).

\bibitem{st3} P.\ Jacobs, STAR Collaboration, talk given at HIC 03,
McGill University (June 2003).

\bibitem{star3} J.\ Adams {\it et al.}, STAR Collaboration,
nucl-ex/0310004.

\bibitem{hy4} R.\ C.\ Hwa, and C.\ B.\ Yang, Phys.\ Rev.\ C {\bf 66},
025204 (2002).

\bibitem{adl} S.\ S.\ Adler, PHENIX Collaboration, Phys.\ Rev.\
Lett.\  {\bf 91}, 172301 (2003).

\bibitem{star2} J.\ Adams {\it et al.}, STAR Collaboration,  Phys.\
Rev.\ Lett.\  {\bf 91}, 072304 (2003).

\end{thebibliography}
\end{document}